 \documentclass[preprint]{aastex}
\def\simgr{\,\hbox{\hbox{$ > $}\kern -0.8em \lower 1.0ex\hbox{$\sim$}}\,}
\def\simle{\,\hbox{\hbox{$ < $}\kern -0.8em \lower 1.0ex\hbox{$\sim$}}\,}

\shortauthors{THORSTENSEN \& PETERS.}
\shorttitle{Longer-Period Cataclysmics}
\begin{document}
\title{Spectroscopy of Four Cataclysmic Variables with Periods above 7 Hours
\footnote{Based on observations obtained at the MDM Observatory, operated by
Dartmouth College, Columbia University, Ohio State University, and
the University of Michigan.}
}

\author{Christopher S. Peters and John R. Thorstensen}
\affil{Department of Physics and Astronomy\\
6127 Wilder Laboratory, Dartmouth College\\
Hanover, NH 03755-3528;\\
christopher.s.peters@dartmouth.edu\\
john.thorstensen@dartmouth.edu}

\begin{abstract}

We present spectroscopy of four cataclysmic variables.  Using radial velocity 
measurements, we find orbital periods $P_{\rm orb}$ for the first time. The 
stars and their periods are GY Hya, 
0.347230(9) d; SDSS J204448-045929, 1.68(1) d; V392 Hya, 
0.324952(5) d; and RX J1951.7+3716, 0.492(1) d. We also detect the 
spectra of the secondary stars, estimate their spectral types, and derive 
distances based on surface brightness and Roche lobe constraints.

\end{abstract}

\section{Introduction}
Cataclysmic variables (CVs) are close binary star systems comprised of a white 
dwarf (the primary) that accretes matter from a less compact companion star
(the secondary) in Roche lobe overflow.  There exist many types of CVs, 
providing unique laboratories in which to study various aspects of 
astrophysics.  \citet{war95} gives an overview of the field.

The most fundamental characteristic of a CV is the period of its 
orbit, $P_{\rm orb}$.  When an eclipse is evident, 
the period is incontrovertible.  In non-eclipsing CVs, the period may be found 
using radial velocities.  Often, a spectral contribution from the 
secondary star is present.  This provides useful clues to the distance and 
the evolutionary state of the system.
   
In the current study, we discuss four CVs:  GY Hya, SDSS J204448-045929, V392 
Hya, and RXJ1951.7+3716.  None of them have extensive spectroscopic studies 
in the literature.  We report their basic properties (period, spectral type 
of secondary) and infer distances.  All four stars have periods 
above 7 hours, the longest period being that of SDSS J204448-045929, 
$P_{\rm orb} = 1.68(1)$ d.  As stated in \citet{thfe04} (hereafter TFT04), 
longer-period systems provide information on CV evolution; in particular, 
many systems appear to have begun nuclear evolution prior to mass 
transfer \citep{bk00}.  TFT04 provide a background in long-period CV analysis. 

Section 2 describes the observations and analysis, and \S 3 gives results.  
In \S 4, we provide a brief discussion.

\section{Techniques}

Table~\ref{tab:journal} lists the observations, all of which were taken with 
the 2.4-meter Hiltner telescope at MDM Observatory in Kitt Peak, Arizona.  We 
used the 
`modular' spectrograph, a 600 line mm$^{-1}$ grating, a 1-arcsec slit, and 
a SITe 2048$^2$ CCD detector yielding 2 \AA\ pixel$^{-1}$ from 4210 to 7560 
\AA\ (vignetting severely toward the ends), and typical resolution of 3.5 
\AA\ FWHM.  On most observing runs, spectra of comparison lamps were taken 
whenever the telescope
was moved to achieve accurate wavelength calibration.  On some observing runs, 
comparison lamp spectra were not taken during the night, and the wavelength 
solution derived from lamps was shifted by an amount determined from the 
night-sky lines.  For a detailed discussion of 
calibration procedures, see TFT04.

\subsection{Period Determination}
Our main objective was to find orbital periods.  To
this end, we measured radial velocities in the spectra using both
absorption and emission lines, when possible.  The absorption lines originate 
in the atmosphere of the 
secondary star.  Absorption line velocities were found 
using the IRAF cross-correlation radial velocity package {\it xcsao} 
\citep{kurtzmink}.  The routine was run on the wavelength range 6000 to 6500 
\AA.  Uncertainties, based on the $R$-statistic of \citet{tondav}, are typically 
less than 10 km s$^{-1}$.  For a handful of low signal-to-noise exposures, the 
cross-correlations were not formally significant, leading to unphysical
velocities which were excluded from further analyses.

Emission line velocities were found by measuring the shift of H$\alpha$
via the convolution technique of \citet{sy80} and \citet{shafter83}.  This 
method consists of convolving two antisymmetric Gaussians with the emission 
line, with an adjustable separation parameter, $\alpha$, and searching for the 
zero of the convolution.  Typical values of $\alpha$ are
12-15 \AA.  The idea (valid or not) is that the two Gaussians measure the wings of the 
line, which may arise in a symmetrical portion of the accretion-disk 
emission close to the white dwarf.  The best parameters for the convolution 
function are found by searching for the largest value of $\alpha$ which yields 
reliable results.  This was done for all stars, 
regardless of line profile (single- or double-peaked).

We ran a period-search 
algorithm which fit general least-squares sinusoids of the form 
$$v(t) = A\cos(\omega t) + B\sin(\omega t) + C$$ to the time series with a 
range of equally-spaced frequencies, $\omega$.  Ideally, a periodogram of 1/$\chi^2$ 
versus frequency shows a lone peak corresponding to the true frequency.  
Due to the uneven sampling of the time-series, alias (false) periods often 
appear in the periodogram.  We used a Monte Carlo algorithm developed by 
\citet{thfreed85} to assess confidence with which the highest peak in the 
periodogram can be identified with the true orbital frequency.  Once the 
optimal period was found, we fit the time-series with sinusoids of the 
form $$v(t) = \gamma + K\sin[2 \pi(t - T_0)/P]$$ using a hybrid linear 
least-square algorithm.  This procedure is 
described in detail in TFT04.  

\begin{figure}
\epsscale{0.8}
\plotone{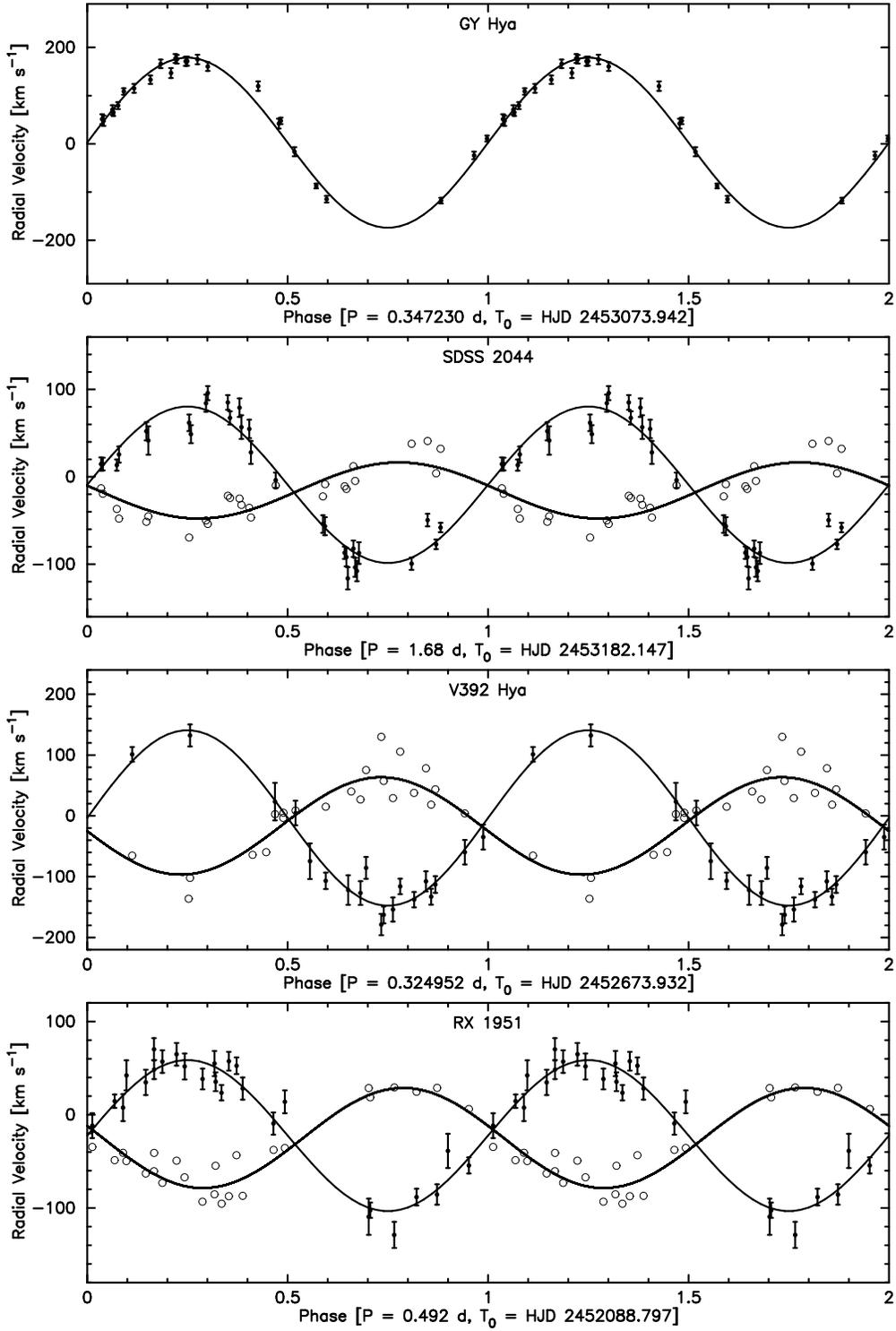}
\caption{Absorption (solid dots with error bars) and emission (open
circles) radial velocities folded on the adopted orbital periods.
Best-fit sinusoids are superposed.  All data are shown twice
for continuity.  For GY Hya, 
the curve is folded about the only available period, that of the absorption 
line velocities.
}
\label{fig:folpl}
\end{figure}

In Figure~\ref{fig:folpl}, radial velocity curves are shown, 
folded twice for continuity.  When both absorption and emission line 
velocities are available, the radial velocity curves are folded using the 
weighted mean period.  The parameters for the least-squares best-fit curves are 
given in Table~\ref{tab:parameters}.  The absorption and emission line curves 
are very close to being $\frac{1}{2}$ cycle out-of-phase.  
If the curves are consistent with $\Delta \phi = 0.5$, the mass ratio of the 
secondary to the primary may be estimated to be $q = M_2/M_1= K_1/K_2$.  We 
caution the reader that the 
measured $K$ values may misrepresent the orbital motion of the components, 
especially $K_1$ which originates from the accretion disk.  Having the 
emission and absorption velocities $\frac{1}{2}$ cycle out-of-phase is a 
necessary, but not sufficient condition for the $K_1$ to represent the 
motion of the white dwarf.

\subsection{Spectral Decomposition}
For spectral decomposition, we prepared an average spectrum for each system 
in which the individual exposures were shifted to the absorption line rest 
frame.  We have a collection of standard K- and M-star 
obtained with the same instrumental setup as the present data.  
These K and M stars were classified by \citet{keenan89} and 
\citet{boe76}, respectively.  The library of template 
spectra were scaled and subsequently subtracted from the averaged 
spectra in order to eliminate the absorption lines due to the secondary.  We 
did not rotationally-broaden our template spectra prior to subtraction 
because the rotational broadening expected for the secondary stars is
small compared to our spectral resolution.  We examined the residual 
spectra by eye.  In an ideal case, the absorption lines would disappear
entirely, leaving only the continuum and emission lines, but in reality
the subtractions were never perfect.  Nonetheless, the subtractions were 
fairly good.  In all the cases considered here, the spectral type of the 
secondary could be estimated to $\pm 1$ subclass, and the secondary's 
contribution to the flux could be estimated to $\pm 15\%$.

Using the known periods and spectral types, inferences about the systems have 
been made based on physical considerations and empirical relations.  We assumed 
that the secondaries fill their Roche lobes, and are undergoing mass transfer.  
Note carefully that we {\it did not} assume that the 
secondaries follow a main-sequence spectral type versus absolute 
magnitude (sp-$M_V$) relationship.  
Instead, we used a relationship between the surface brightness and the 
spectral type for normal stars derived by \citet{beu04}.  To convert
the surface brightness to $M_V$, we of course need a secondary-star radius 
$R_2$.  This is strongly constrained by the Roche geometry and the orbital 
period. To estimate $R_2$ we begin with an analytical approximation
to the Roche lobe radius, given by \citet{beu98}, which shows that,
at a given $P_{\rm orb}$, $R_2$ is almost independent of the primary
star mass $M_1$, and is dependent only on the cube root of $M_2$.  We have
no direct measure of $M_2$, but we can estimate a range of plausible values
using the evolutionary scenarios of \citet{bk00} as a guideline, at
least for systems with $P_{\rm orb} < 10$ h.
Armed with our estimate of the secondary's radius, we transformed 
the surface brightness to $M_V$.  The synthetic 
apparent magnitude of the secondary, $m_V$, was found from the scaled 
template spectra used during spectral decomposition using the 
IRAF task {\it sbands} and the \citet{bessell} tabulation of the 
$V$ passband.  Finally, visual extinction values, $A_V$, were estimated 
from infrared dust maps of \citet{schlegel98}, and $A_V$, $m_V$, 
and $M_V$ were combined to yield a distance.  The dust extinction
estimate is imperfect, because (a) the ISM can be patchy on scales
smaller than the 6-arcmin resolution of the maps, and (b) the 
maps give estimates of the total extinction to the edge of the 
Galaxy, rather than to the star of interest.  However, 
for three of the four objects studied here, the
Galactic latitudes are high enough that the uncertain 
extinction should have a fairly minor effect on the distance.

\section{The Individual Stars}

Our results are summarized in tables and figures.
Table~\ref{tab:line_features} shows measurements of spectral line
properties based on flux-averaged spectra.  All radial velocity
measurements included in the analysis are listed in
Table~\ref{tab:velocities}.  Table~\ref{tab:parameters} lists
parameters of the best-fit sinusoids to the velocity time series.
Table~\ref{tab:inferences} shows derived characteristics of the
secondary and inferred distances.  Fig.~\ref{fig:folpl} shows folded
radial velocity curves, Fig. ~\ref{fig:v392hya_comp} displays the
spectrum of V392 Hya during two separate observing runs, and
Fig.~\ref{fig:subspec} shows the flux-averaged spectra of the four
objects before and after spectral decomposition.
 
\subsection{GY Hya}

\citet{dow01} list GY Hya as an unconfirmed U Gem star, a subclass of
dwarf nova (DN).  \citet{zwitter} obtained a spectrum; strong emission
from H$\alpha$, H$\beta$, and HeII were present.  Our spectroscopic
observations span a baseline of 111 days.  Our spectrum
(Fig.~\ref{fig:subspec}) shows weaker emission features and strong
absorption lines, indicating a significant contribution from the
secondary.  The absorption line velocities have large amplitude ($K_2$
$\sim 180\ $km s$^{-1}$) and small scatter about a best-fit sinusoid
of period, $P_{\rm orb} = 0.347230(9)$ d, or 500 min.  A Monte Carlo
simulation yields a discriminatory power of 98$\%$ for this $P_{\rm
orb}$.  Time-series photometry from \citet{monard} shows an eclipse
with a period of  $P_{\rm orb} = 0.347237(1)$ d;  our period is
consistent with this.  The emission lines were weak, thus a reliable
fit to the H$\alpha$ emission line velocities was not found and not
included in Fig.~\ref{fig:folpl}.

Spectral decomposition of our flux-averaged spectrum shows a secondary
of type K4 or K5.  The synthetic V magnitude of the secondary based on
the decomposition is $16.7 \pm 0.2$.  This uncertainty includes both
the uncertainty in spectral type and also in the factors used to
multiply the template spectra.  The $P_{\rm orb}$ falls within the
range of evolutionary scenarios considered by \citet{bk00}, so for
purposes of defining the Roche lobe, we estimate $M_2 = 0.65 \pm 0.10$
$\rm M_{\odot}$.  This corresponds to $R_2 = 0.84 \pm 0.05$ $\rm
R_{\odot}$ at this $P_{\rm orb}$, which using surface brightnesses
from \citet{beu04} in turn implies $M_V = 5.9 \pm 0.2$.  Assuming an
extinction $A_V = 0.3$, this yielded a distance of
$1260^{+190}_{-160}$ pc.  

With no reliable emission line velocities, an estimate of the mass ratio 
could not be made.  The fact that GY Hya is an eclipsing system constrains 
the inclination to be greater than $\sim$ 70 degrees.  The available evidence 
does not suggest that the masses are unusual; for example, with $q \le 1$ and 
$M_2 = 0.65 \pm 0.10$ $\rm M_{\odot}$, there is a comfortable fit with 
$i = 80$ degrees at $M_1 = 0.7$ $\rm M_{\odot}$ and 
$M_2 = 0.65$ $\rm M_{\odot}$.

\subsection{SDSS J204448-045929}

The subclass of SDSS J204448-045929 (SDSS 2044, for brevity) is
unknown, as it was recently discovered as a CV in the Sloan Digital
Sky Survey \citep{szk03}.  Our data span 10 days and had high
signal-to-noise (S/N) ratios.  The spectra showed intrinsically strong
and narrow emission lines, most notably H$\alpha$.  Both the
absorption and emission line velocities are measureable, although
amplitudes are low.  The orbital period, $P_{\rm orb} = 1.68(1)$ d, is
unusually long for a CV.  According to the {\it The Catalog and Atlas
of CVs: The Living Edition} \citep{dow01}, only 7$\%$ of CVs with
known periods (440 total, including SDSS 2044) have periods above 9
hours.  However, a 97$\%$ Monte Carlo discriminatory power shows that
our preferred period is very likely to be correct.

We find that the absorption and emission line radial velocity curves
in Figure~\ref{fig:folpl} are consistent with $\Delta \phi = 0.5$.
This allows us to estimate a mass ratio, $q = 0.36^{+0.13}_{-0.12}$ .
Given that the mass of the white dwarf needs to be less than the
Chandrasekhar limit of 1.4 $\rm M_{\odot}$, this requires that the
secondary have a mass $\le 0.7$ $\rm M_{\odot}$.  The period of SDSS
2044 ($\sim 40$ h) is well above the 10.0 h upper limit of the
evolutionary scenarios of \citet{bk00}, so we could not use their
results.  Given the above mass ratio constraints, we adopt a
conservative range of mass for the secondary to be $M_2 = 0.4 \pm 0.3$
$\rm M_{\odot}$.  The secondary's radius is dependent upon the
cube-root of the mass, so the although the uncertainty in mass is
large, the radius uncertainty is not:  $R_2 = 1.9 \pm 0.5$ $\rm
R_{\odot}$ at $P_{\rm orb} = 1.68(1)$ d.  Given the range in $q$ and a
conservative range in white dwarf mass of $M_1 = 0.9 \pm 0.4$ $\rm
M_{\odot}$, the inclination is within $45 \pm 10$ degrees.  

A clean spectral decomposition gives a secondary spectral type of K4
or K5, with corresponding synthetic V apparent magnitude of $18.4 \pm
0.2$.  Using the radius for the secondary, the surface brightness is
transformed into an absolute magnitude of $M_V = 7.5 \pm 0.7$.  With
an extinction of $A_V = 0.2$, we find a distance of
$1380^{+530}_{-380}$ pc.

\subsection{V392 Hya}

Similar to GY Hya, V392 Hya is also listed by \citet{dow01} as an
unconfirmed U Gem star (DN).  V392 Hya, or EC 10565-2858, was
discovered in the Edinburgh-Cape Blue Object Survey.  \citet{chen}
show a spectrum that spans $\lambda$$\lambda$ 3800-5000 \AA; our
spectral range is 4210-7560\AA.  Therefore, comparison is limited; the
only spectral feature in common is H$\beta$ emission.  \citet{chen}
report two periods, $\sim$ 8 h and $\sim$ 12 h, derived from
time-resolved spectroscopy.  Due to gaps in their time series, they
were unable to distinguish between the two periods.  Our time series
is comprised of a lone observation in 2002, followed a year later by 4
consecutive nights, and finally 6 nights scattered through 2004.

\begin{figure}
\epsscale{0.95}
\plotone{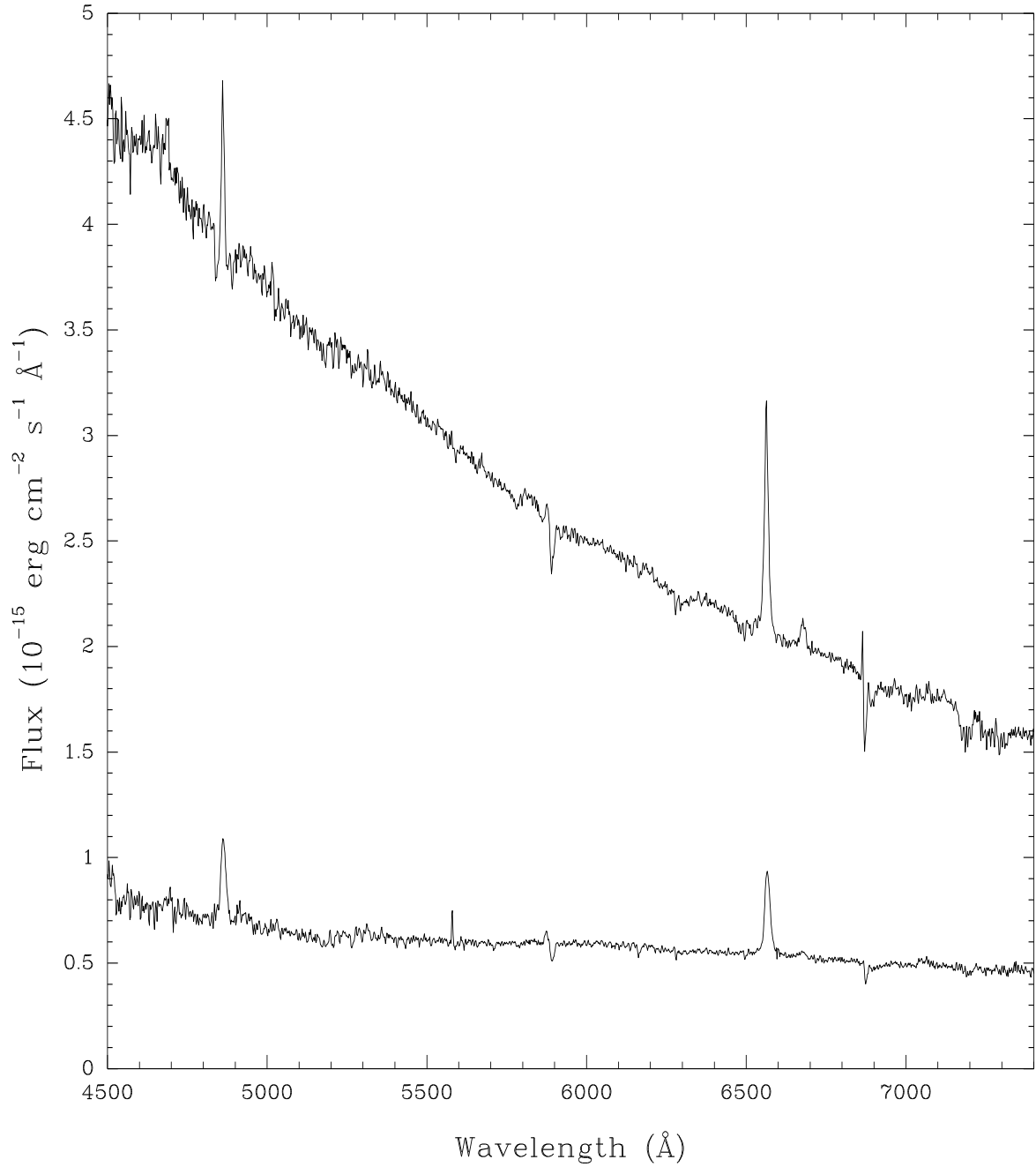}
\caption{Mean spectra of V392 Hya in different states.  Top/bottom spectrum 
was taken in 2004/2003, respectively.  Note the differences in mean flux 
level, continuum shape, and line strength.
}

\label{fig:v392hya_comp}
\end{figure}

Figure~\ref{fig:v392hya_comp} shows V392 Hya in two different states
during January 2004 and February 2003.  This is evident from the
continua.  The mean flux level differs by a factor of $\sim 4$.
Differences in spectral features are seen, most notably the strength
of H$\alpha$ emission and the structure of the H$\beta$ line.  As
shown in Table~\ref{tab:line_features} and
Figure~\ref{fig:v392hya_comp}, the H$\alpha$ line becomes thinner and
stronger by a factor of $\sim 2$ in outburst.  The H$\beta$ line
changes from emission superposed on flat continuum during the lower
state to emission superposed on a wider absorption line during the
higher state.  This is commonly thought to be due to the optically
thick accretion disk. 

Despite the different states, the absorption line velocities from the
late-type secondary are measureable in all our data and yield a
best-fit period of $P_{\rm orb} = 0.324952(5)$ d, or 468 min.  A Monte
Carlo simulation to discriminate between alias periods returns a
confidence level of $\sim 90\%$, which is the lowest for the stars
presented in this paper.  Emission line velocities based on H$\alpha$
had more scatter about the best-fit curve than the absorption line
velocities, but fit reasonably well, as seen in
Figure~\ref{fig:folpl}.  The radial velocity curves for V392 Hya are
also consistent with $\Delta \phi = 0.5$, leading to an estimate of $q
= 0.55 \pm 0.15$.  This constrains the inclination to be $i = 45 \pm
13$ degrees, assuming a conservative white dwarf mass of $M_1 = 0.9
\pm 0.4$ $\rm M_{\odot}$.

Spectral decomposition was done using the observations in quiescence
since the absorption lines (due to the secondary star) were more
prominent.  Visual inspection led to an inferred secondary of spectral
type K5 or K6, with corresponding synthetic V magnitude of $m_V = 18.9
\pm 0.2$.  A K5/6 secondary with a period of $\sim 7.8$ h implied
(using \citet{bk00}) a mass and radius of $0.60 \pm 0.05$ $\rm
M_{\odot}$ and $0.78 \pm 0.03$ $\rm R_{\odot}$, respectively.  This
radius translated the surface brightness calculation of \citet{beu04}
into an absolute magnitude of $M_V = 6.0 \pm 0.2$, which, when
combined with an extinction of 0.2 magnitudes and $m_V$, gave an
inferred distance of $3470^{+510}_{-450}$ pc.

\subsection{RXJ1951.7+3716}

\begin{figure}
\epsscale{0.88}
\plotone{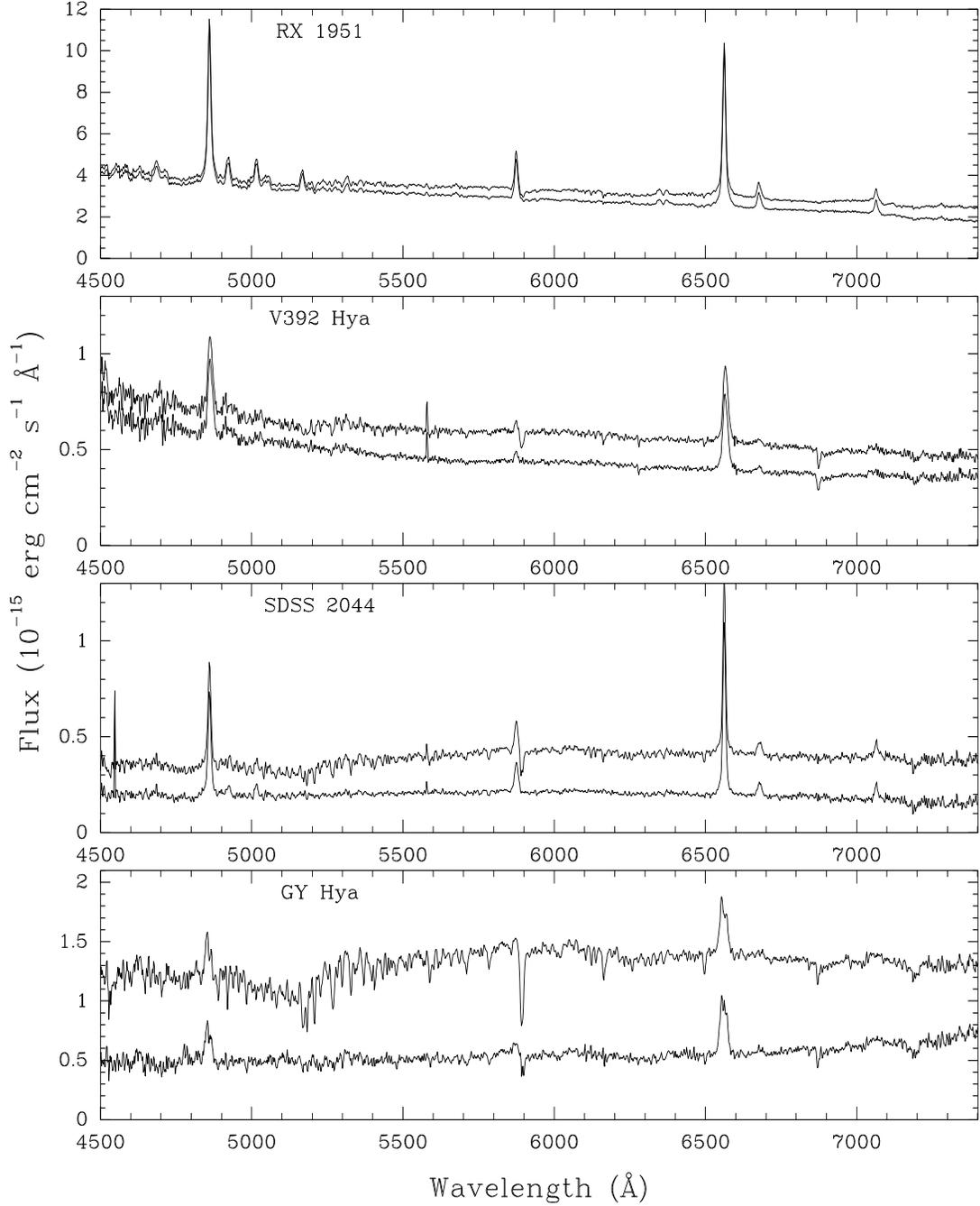}
\caption{A montage of spectra.  The vertical scale in each plot is in units
of $10^{-15}$ erg s$^{-1}$ cm$^{-2}$ \AA $^{-1}$, subject to 
calibration uncertainties of some tens of percent.  The 
lower trace in each panel shows the data after a scaled
late-type star has been subtracted away (see text and 
Table~\ref{tab:inferences}).  In all cases the original spectra are 
shifted into the rest frame of the secondary star before averaging. 
}
\label{fig:subspec}
\end{figure}

\citet{motch98} classify RXJ1951.7+3716 (RX 1951, hereafter) as a CV,
with no specified subclass.  Our data span 10 days in 2001, with
coverage over an 8-hour range in hour angle to discriminate the daily
cycle alias.  The mean flux level for the spectra is $\ge 2.5$ times
higher than the other three objects in this paper.  In
Figure~\ref{fig:subspec}, the small difference between the upper and
lower spectra shows the weak, but present, contribution from the
secondary.  

The emission lines (particularly H$\alpha$ and H$\beta$) are strong
and narrow, yielding small uncertainty in emission line velocities
(see Table~\ref{tab:velocities}).  Despite the weakness of the
secondary contribution, absorption line velocities are measurable.
The weighted mean period found is $P_{\rm orb} = 0.492(1)$ d.  Monte
Carlo simulations rule out other alias periods with a discriminatory
power of $\ge 90 \%$.  With a small $K_2$ of $81 \pm 7$ km s$^{-1}$,
the inclination is expected to be low.  We find that for a white dwarf
mass range of $0.9 \pm 0.4$ $\rm M_{\odot}$ and $q \le 1$, $i = 25 \pm
10$ degrees.

The spectral type of the secondary is estimated to be K7.5 or M0.5;
the lower spectrum in Figure~\ref{fig:subspec} is after an M0.5 star
has been subtracted.  The synthetic magnitude calculated for the
secondary star is $m_V = 17.6 \pm 0.2$.  A secondary mass of $0.6 \pm
0.3$ $\rm M_{\odot}$ is extrapolated from \citet{bk00} since the
period lies slightly above the cutoff of 10.0 hours.  At $P_{\rm orb}
= 0.492(1)$ d, this mass range corresponds to a radius of $R_2 = 0.99
\pm 0.18$ $\rm R_{\odot}$.  \citet{beu04} surface brightnesses were
converted into an absolute magnitude of $M_V = 7.4 \pm 0.6$.  The
extinction value used is large ($A_V = 0.8$) due to the fact that RX
1951 is only $\sim 5$ degrees from the Galactic plane.  We find RX
1951 to be at a distance of $760^{+240}_{-180}$ pc.  It is not
surprising to find that this source is closer than the other three due
to its greater apparent brightness.

\section{Discussion}

We find orbital periods for these four systems to be above 7 hours;
the period of SDSS 2044, $P_{\rm orb} = 1.68(1)$ d is notably long for
a CV.  In longer-period systems, the secondaries are usually cooler
than expected for main-sequence stars filling their Roche lobes at the
observed period \citep{beu98}.  Referring to the plot of spectral type
vs. period in \citet{bk00}, it is clear that the four secondary stars
in this paper do not lie on zero-age main sequence (ZAMS) evolutionary
tracks, but rather are significantly cooler, most likely due to
commencement of nuclear evolution prior to the onset of mass transfer
\citep{bk00}.  Although the models of \citet{bk00} do not apply to
periods above 10 hours, and thus may not be directly applied to SDSS
2044 and RX 1951, these two systems are the furthest from the ZAMS.

The systems in this paper have been selected based on their longer
periods.  Therefore, they do not represent the full range in CV
periods.  For example, dwarf novae with hydrogen-rich secondaries have
periods as short as 75 minutes.  A goal of studying CVs is to obtain
an accurate catalog of their population.  To do so, different methods
of CV discovery are essential, as is evident from the subset presently
studied.  V392 Hya was identified in the Edinburgh-Cape Blue Object
Survey, while SDSS 2044 and RX 1951 were discovered by SDSS and ROSAT,
respectively.  The wide range in techniques and energies is beneficial
for finding these systems. 

We find distances to the CVs via spectroscopic parallax.  This method
relies on surface brightness calculations of the secondary star
filling its Roche lobe, but more importantly upon accurate
identification of the secondary's spectral type.  We are able to
detect and deduce the spectral type (via spectral decomposition) of
the secondary star to within one subclass.  This is sufficient to
yield distances with 15$\%$ uncertainty in favorable cases.

{\it Acknowledgements}

We gratefully acknowledge support from NSF grants AST-9987334 and 
AST-0307413.  Also, we thank Bill 
Fenton for taking some of the data, and the MDM staff for observing 
assistance.

\clearpage

\begin{deluxetable}{lrcc}
\tablewidth{0pt}
\tablecolumns{4}
\tablecaption{Journal of Observations}
\tablehead{
\colhead{Date} &
\colhead{$N$} &
\colhead{HA (start)}  &
\colhead{HA (end)} \\
\colhead{[UT]}  &
 &
\colhead{[hh:mm]} &
\colhead{[hh:mm]} \\
}
\startdata

\cutinhead{GY Hya:} 
2004 Mar 7 &  8 & $ -2:05$ & $ +1:36$ \\ 
2004 Mar 9 &  9 & $ -0:33$ & $ +1:29$ \\ 
2004 Jun 22 &  3 & $ -0:15$ & $ +2:21$ \\ 
2004 Jun 25 &  2 & $ +0:02$ & $ +0:15$ \\ 
2004 Jun 26 &  1 & $ -0:06$ & $ -0:06$ \\ 
2004 Jul 1 &  2 & $ +0:18$ & $ +0:35$ \\
\cutinhead{SDSS 2044-04:}
2004 Jun 22 &  2 & $ -1:58$ & $ -1:47$ \\ 
2004 Jun 23 &  3 & $ -1:24$ & $ -1:05$ \\ 
2004 Jun 24 &  4 & $ -0:39$ & $ +1:14$ \\ 
2004 Jun 25 &  3 & $ -2:13$ & $ +0:43$ \\ 
2004 Jun 26 &  5 & $ -3:08$ & $ +0:30$ \\ 
2004 Jun 27 &  2 & $ +0:55$ & $ +1:07$ \\ 
2004 Jun 28 &  6 & $ -2:19$ & $ +1:20$ \\ 
2004 Jun 29 &  2 & $ -3:41$ & $ -3:29$ \\ 
2004 Jun 30 &  1 & $ +1:33$ & $ +1:33$ \\ 
2004 Jul 1 &  2 & $ -3:02$ & $ -2:49$ \\
\cutinhead{V392 Hya:}
2002 Jan 22 &  2 & $ +0:07$ & $ +0:18$ \\ 
2003 Jan 31 &  1 & $ +0:17$ & $ +0:17$ \\ 
2003 Feb 1 &  3 & $ -1:21$ & $ +2:42$ \\ 
2003 Feb 2 &  2 & $ -0:10$ & $ +1:11$ \\ 
2003 Feb 3 &  5 & $ -2:44$ & $ -1:07$ \\ 
2004 Jan 13 &  1 & $ -1:23$ & $ -1:23$ \\ 
2004 Jan 16 &  1 & $ -0:58$ & $ -0:58$ \\ 
2004 Jan 17 &  1 & $ +2:12$ & $ +2:12$ \\ 
2004 Jan 18 &  1 & $ -0:47$ & $ -0:47$ \\ 
2004 Jan 19 &  1 & $ +2:47$ & $ +2:47$ \\ 
2004 Mar 7 &  1 & $ +1:58$ & $ +1:58$ \\
\cutinhead{RX 1951:}
2001 Jun 23 &  2 & $ +0:52$ & $ +0:57$ \\ 
2001 Jun 24 &  1 & $ +2:07$ & $ +2:07$ \\ 
2001 Jun 25 &  3 & $ -2:02$ & $ +2:02$ \\ 
2001 Jun 26 &  5 & $ -4:38$ & $ +2:15$ \\ 
2001 Jun 27 &  3 & $ -2:03$ & $ +2:19$ \\ 
2001 Jun 28 &  4 & $ -5:19$ & $ +2:37$ \\ 
2001 Jun 29 &  2 & $ -1:19$ & $ +2:28$ \\ 
2001 Jul 1 &  4 & $ -4:52$ & $ +2:44$ \\ 
2001 Jul 2 &  2 & $ +1:06$ & $ +2:45$ \\
\enddata
\label{tab:journal}
\end{deluxetable}

\clearpage

\begin{deluxetable}{lrcc}
\tablewidth{0pt}
\tablecolumns{4}
\tablecaption{Emission Features}
\tablehead{
\colhead{Feature} &
\colhead{E.W.\tablenotemark{a}} &
\colhead{Flux\tablenotemark{b}}  &
\colhead{FWHM \tablenotemark{c}} \\
 &
\colhead{(\AA )} &
\colhead{(10$^{-16}$ erg cm$^{-2}$ s$^{1}$)} &
\colhead{(\AA)} \\
}
\startdata
\cutinhead{GY Hya:} 
  H$\beta$  & $ 7$ & $86$ & 22 \\ 
  NaD & $ -6$ & $-93$ & 13 \\ 
  H$\alpha$  & $ 9$ & $122$ & 28 \\ 
\cutinhead{SDSS 2044-04:}
           H$\gamma$ & $ 28$ & $ 84$ & 17 \\
  HeI $\lambda 4471$ & $  7$ & $ 22$ & 13 \\
            H$\beta$ & $ 17$ & $ 64$ & 12 \\
  HeI $\lambda 5015$ & $  3$ & $ 9$ & 13 \\
  HeI $\lambda 5876$ & $  6$ & $ 26$ & 14 \\
           H$\alpha$ & $ 21$ & $ 95$ & 10 \\
  HeI $\lambda 6678$ & $  3$ & $  13$ & 18 \\
  HeI $\lambda 7067$ & $  4$ & $  15$ & 19 \\
\cutinhead{V392 Hya (high state):}
           H$\gamma$ & $ 1$ & $53$ & 6 \\ 
  HeII $\lambda 4686$ & $ 1$ & $ 25$ & 11 \\
            H$\beta$ & $ 3$ & $106$ & 12 \\ 
  HeI $\lambda 5876$ & $ 1$ & $ 8$ & 9 \\ 
  NaD & $ -1$ & $ -29$ & 13 \\
           H$\alpha$ & $ 7$ & $155$ & 15 \\ 
  HeI $\lambda 6678$ & $  1$ & $ 21$ & 17 \\
\cutinhead{V392 Hya (low state):}
            H$\beta$ & $ 12$ & $80$ & 20 \\ 
  HeI $\lambda 5876$ & $ 1$ & $ 7$ & 11 \\ 
  NaD & $ -2$ & $ -11$ & 12 \\
           H$\alpha$ & $ 14$ & $77$ & 19 \\ 
  HeI $\lambda 6678$ & $  2$ & $ 8$ & 28 \\
\cutinhead{RX 1951:}
           H$\gamma$ & $ 21$ & $945$ & $ 13$ \\ 
   HeI $\lambda 4471$ & $ 9$ & $378$ & 18 \\
  HeII $\lambda 4686$ & $3$ & $113$ & 17 \\
            H$\beta$ & $ 26$ & $1010$ & 13 \\ 
   HeI $\lambda 4921$ & $ 4$ & $169$ & 15 \\
   HeI $\lambda 5015$ & $ 3$ & $131$ & 12 \\
   Fe $\lambda 5169$ & $ 3$ & $115$ & 13 \\
  HeI $\lambda 5876$ & $ 7$ & $ 237$ & 12 \\ 
  NaD & $ -1$ & $ -38$ & 12 \\
           H$\alpha$ & $ 29$ & $922$ & 13 \\ 
  HeI $\lambda 6678$ & $  4$ & $ 110$ & 15 \\	
  HeI $\lambda 7067$ & $  4$ & $ 108$ & 17 \\
\enddata
\tablenotetext{a}{Emission equivalent widths are counted as positive.}
\tablenotetext{b}{Absolute line fluxes are uncertain by a factor of about
2, but relative fluxes of strong lines
are estimated accurate to $\sim 10$ per cent.}
\tablenotetext{c}{From Gaussian fits.}
\label{tab:line_features}
\end{deluxetable}

\clearpage

\begin{deluxetable}{lrcrc}
\tablewidth{0pt}
\tablecolumns{5}
\tablecaption{Radial Velocities}
\tablehead{
\colhead{Modified JD} &
\colhead{$v_{\rm abs}$} &
\colhead{$\sigma_{v_{\rm abs}}$}  &
\colhead{$v_{\rm emn}$} &
\colhead{$\sigma_{v_{\rm emn}}$} \\

}
\startdata
\cutinhead{GY Hya}
53071.8712  & $   51$ & $  10$ &  \nodata & \nodata \\
53071.8802  & $   70$ & $  10$ &  \nodata & \nodata \\
53071.9361  & $  175$ & $   9$ &  \nodata & \nodata \\
53071.9450  & $  171$ & $   9$ &  \nodata & \nodata \\
53071.9539  & $  175$ & $  10$ &  \nodata & \nodata \\
53071.9628  & $  160$ & $   9$ &  \nodata & \nodata \\
53072.0064  & $  120$ & $  10$ &  \nodata & \nodata \\
53072.0243  & $   42$ & $  10$ &  \nodata & \nodata \\
53073.9296  & $  -24$ & $   8$ &  \nodata & \nodata \\
53073.9408  & $   11$ & $   6$ &  \nodata & \nodata \\
53073.9559  & $   48$ & $  11$ &  \nodata & \nodata \\
53073.9644  & $   65$ & $   8$ &  \nodata & \nodata \\
53073.9734  & $  109$ & $   7$ &  \nodata & \nodata \\
53073.9823  & $  115$ & $   9$ &  \nodata & \nodata \\
53073.9965  & $  133$ & $   9$ &  \nodata & \nodata \\
53074.0054  & $  166$ & $   9$ &  \nodata & \nodata \\
53074.0143  & $  147$ & $  10$ &  \nodata & \nodata \\
53178.6563  & $  -87$ & $   5$ &  \nodata & \nodata \\
53178.6653  & $ -115$ & $   7$ &  \nodata & \nodata \\
53178.7645  & $ -118$ & $   6$ &  \nodata & \nodata \\
53181.6594  & $  177$ & $   9$ &  \nodata & \nodata \\
53181.6683  & $  171$ & $   9$ &  \nodata & \nodata \\
53182.6514  & $   79$ & $   7$ &  \nodata & \nodata \\
53187.6536  & $   48$ & $   7$ &  \nodata & \nodata \\
53187.6655  & $  -16$ & $   9$ &  \nodata & \nodata \\

\cutinhead{SDSS2044:}
53178.8446  & $   15$ & $   8$ & $  -13$ & $   4$ \\
53178.8521  & $   15$ & $   7$ & $  -19$ & $   4$ \\
53179.8652  & $  -87$ & $   7$ & $  -11$ & $   4$ \\
53179.8728  & $  -92$ & $  11$ & $  -14$ & $   4$ \\
53179.8786  & $ -116$ & $  13$ &  \nodata & \nodata \\
53180.8940  & $   62$ & $  10$ & $  -70$ & $   5$ \\
53180.9015  & $   49$ & $  10$ &  \nodata & \nodata \\
53180.9645  & $   84$ & $  10$ & $  -50$ & $   5$ \\
53180.9720  & $   96$ & $   8$ & $  -54$ & $   5$ \\
53181.8260  & $  -99$ & $   7$ & $   38$ & $   5$ \\
53181.8934  & $  -50$ & $   7$ & $   41$ & $   5$ \\
53181.9477  & $  -58$ & $   5$ & $   32$ & $   5$ \\
53182.7852  & $   79$ & $  11$ & $  -25$ & $   6$ \\
53182.7928  & $   57$ & $  14$ & $  -32$ & $   6$ \\
53182.8258  & $   55$ & $  11$ & $  -36$ & $   6$ \\
53182.8330  & $   28$ & $  13$ & $  -47$ & $   7$ \\
53182.9364  & $   -4$ & $   9$ & $  -10$ & $   5$ \\
53183.9507  & $   13$ & $   6$ & $  -37$ & $   5$ \\
53183.9596  & $   26$ & $   9$ & $  -48$ & $   5$ \\
53184.8143  & $  -54$ & $  10$ & $  -22$ & $   5$ \\
53184.8232  & $  -57$ & $  10$ & $   -9$ & $   6$ \\
53184.9419  & $  -82$ & $   9$ & $   12$ & $   6$ \\
53184.9495  & $ -104$ & $  11$ & $   -5$ & $   6$ \\
53184.9569  & $ -108$ & $  12$ &  \nodata & \nodata \\
53184.9656  & $  -87$ & $  12$ &  \nodata & \nodata \\
53185.7547  & $   52$ & $  10$ & $  -51$ & $   6$ \\
53185.7628  & $   42$ & $  16$ & $  -45$ & $   6$ \\
53186.9690  & $  -77$ & $   5$ & $    4$ & $   5$ \\
53187.7761  & $   85$ & $   8$ & $  -22$ & $   5$ \\
53187.7850  & $   67$ & $   8$ & $  -24$ & $   5$ \\
\cutinhead{V392 Hya}
52296.9372  & $ -108$ & $  16$ & $   78$ & $  18$ \\
52296.9448  & $ -113$ & $  14$ & $   44$ & $  11$ \\
52670.9208  & $ -179$ & $  18$ & $  130$ & $  20$ \\
52671.8506  & $ -107$ & $  14$ & $   15$ & $  10$ \\
52671.9360  & $ -133$ & $  13$ & $   18$ & $  10$ \\
52672.0186  & $  101$ & $  12$ & $  -65$ & $  11$ \\
52672.8971  & $ -138$ & $  13$ & $   38$ & $  13$ \\
52672.9531  & $  -35$ & $  20$ &  \nodata & \nodata \\
52673.0392  &  \nodata & \nodata &  $ -136$ & $  22$ \\
52673.7875  & $  -75$ & $  29$ &  \nodata & \nodata \\
52673.8185  & $ -122$ & $  24$ &  \nodata & \nodata \\
52673.8330  & $  -86$ & $  18$ & $   75$ & $  22$ \\
52673.8473  & $ -163$ & $  14$ & $   57$ & $  24$ \\
52673.8548  & $ -154$ & $  20$ & $   29$ & $  21$ \\
53017.8900  &  \nodata & \nodata &  $   -3$ & $  15$ \\
53017.8999  & $    5$ & $  20$ & $    9$ & $  15$ \\
53020.9094  & $ -116$ & $  13$ & $  106$ & $  13$ \\
53022.0388  & $  132$ & $  18$ & $ -102$ & $  16$ \\
53022.9115  & $  -60$ & $  20$ & $    4$ & $  14$ \\
53024.0574  & $   23$ & $  31$ & $    3$ & $   7$ \\
53070.9123  &  \nodata & \nodata &  $   40$ & $  14$ \\
53071.8322  &  \nodata & \nodata &  $    5$ & $  14$ \\
53071.8947  & $ -127$ & $  20$ & $   27$ & $  12$ \\
53073.7569  &  \nodata & \nodata &  $  -64$ & $  10$ \\
53073.7679  &  \nodata & \nodata &  $  -60$ & $   9$ \\
\cutinhead{RX 1951}
52083.9214  & $    8$ & $  14$ & $  -41$ & $   6$ \\
52083.9255  & $   42$ & $  16$ & $  -50$ & $   5$ \\
52084.9711  & $   65$ & $  12$ & $  -49$ & $   4$ \\
52085.7961  & $  -39$ & $  18$ &  \nodata & \nodata \\
52085.9276  & $   49$ & $  10$ & $  -41$ & $   4$ \\
52085.9650  & $   52$ & $  14$ & $  -67$ & $   7$ \\
52086.6848  & $ -102$ & $   8$ & $   19$ & $   4$ \\
52086.7143  & $ -129$ & $  14$ & $   29$ & $   4$ \\
52086.8356  & $  -12$ & $  13$ & $  -34$ & $   6$ \\
52086.9013  & $   35$ & $  14$ & $  -63$ & $   5$ \\
52086.9710  & $   38$ & $  11$ & $  -93$ & $   3$ \\
52087.7898  & $  -54$ & $   9$ & $    6$ & $   4$ \\
52087.8954  & $   70$ & $  12$ & $  -60$ & $   5$ \\
52087.9710  & $   35$ & $  10$ & $  -55$ & $   4$ \\
52088.6510  & $ -109$ & $  19$ & $   29$ & $   9$ \\
52088.7348  & $  -85$ & $  11$ & $   29$ & $   4$ \\
52088.9537  & $   55$ & $  14$ & $  -85$ & $   4$ \\
52088.9807  & $   53$ & $   9$ & $  -43$ & $   4$ \\
52089.8151  & $   15$ & $   7$ & $  -49$ & $   3$ \\
52089.9722  & $   28$ & $  12$ & $  -87$ & $   4$ \\
52091.6617  & $  -88$ & $   9$ & $   25$ & $   7$ \\
52091.8416  & $   57$ & $  12$ & $  -73$ & $   6$ \\
52091.9143  & $   24$ & $   8$ & $  -95$ & $   3$ \\
52091.9780  & $   -9$ & $  12$ & $  -38$ & $   4$ \\
52092.9073  & $   57$ & $  10$ & $  -87$ & $   5$ \\
52092.9760  & $   14$ & $  12$ & $  -36$ & $   7$ \\

\enddata
\label{tab:velocities}
\end{deluxetable}

\clearpage

\begin{deluxetable}{lllrrcc}
\tablecolumns{7}
\footnotesize
\tablewidth{0pt}
\tablecaption{Fits to Radial Velocities}
\tablehead{
\colhead{Data set} & 
\colhead{$T_0$\tablenotemark{a}} & 
\colhead{$P$} &
\colhead{$K$\tablenotemark{b}} & 
\colhead{$\gamma$} & 
\colhead{$N$} &
\colhead{$\sigma$\tablenotemark{c}}  \\ 
\colhead{} & 
\colhead{} &
\colhead{(d)} & 
\colhead{(km s$^{-1}$)} &
\colhead{(km s$^{-1}$)} & 
\colhead{} &
\colhead{(km s$^{-1}$)} \\
}
\startdata
GY Hya (abs) & 53073.942(2) & 0.347230(9) &  176(8) & $ 2(5)$ & 25 &  15 \\
\\[0.5ex]
SDSS 2044 (abs) & 53182.15(2) & 1.68(1) &  90(8) & $-9(5)$ & 30 &  16 \\
SDSS 2044 (emn) & 53183.03(7) & \nodata & 32(8) & $-16(6)$ & 26 &  17 \\
\\[0.5ex]
V392 Hya (abs) & 52673.932(5) & 0.324952(5) &  144(9) & $-3(8)$ & 19 &  23 \\
V392 Hya (emn) & 52673.764(7) & \nodata & 80(14) & $-17(9)$ & 22 &  26 \\
\\[0.5ex]
RX 1951 (abs) & 52088.797(8) & 0.492(1) & 81(7) & $-22(5)$ & 26 & 15 \\
RX 1951 (emn) & 52088.57(1) & \nodata & 54(7) & $-25(6)$ & 25 & 16 \\

\enddata
\tablecomments{Parameters of least-squares sinusoid fits to the radial
velocities, of the form $v(t) = \gamma + K \sin(2 \pi(t - T_0)/P$.
Where both emission and absorption velocities are available, the 
period quoted is the weighted average of the periods derived from 
separated fits to the two data sets, and the period is only given
on the first line.}
\tablenotetext{a}{Heliocentric Julian Date minus 2452000.  The epoch is chosen
to be near the center of the time interval covered by the data, and
within one cycle of an actual observation.}
\tablenotetext{b}{In the text, the absorption and emission line radial velocities 
are denoted $K_2$ and $K_1$, respectively.}
\tablenotetext{c}{Root-mean-square residual of the fit.}
\label{tab:parameters}
\end{deluxetable}

\clearpage

\begin{deluxetable}{llrrrrcr}
\tabletypesize{\scriptsize}
\tablewidth{0pt}
\tablecolumns{8}
\tablecaption{Inferences from Secondary Stars}
\tablehead{
\colhead{Star} &
\colhead{Type} &
\colhead{Synthetic $m_V$}  &
\colhead{Assumed $M_2$\tablenotemark{a}} &
\colhead{Deduced $R_2$} &
\colhead{$M_V$\tablenotemark{b}} &
\colhead{$A_V$} &
\colhead{Distance} \\ 
 &
 & 
\colhead{(mag)} &
\colhead{$\rm M_{\odot}$} &
\colhead{$\rm R_{\odot}$} &
\colhead{(mag)} &
\colhead{(mag)} &
\colhead{(pc)} \\
}
\startdata
GY Hya & K4 or K5 & $16.7 \pm 0.2$ & $0.65 \pm 0.10$ & $0.84 \pm 0.05$
& $5.9 \pm 0.2$ & 0.3 & $1260^{+190}_{-160}$\\
SDSS 2044 & K4 or K5 & $18.4 \pm 0.2$ & $0.4 \pm 0.3$\tablenotemark{c} 
& $1.9 \pm 0.5$ & $7.5 \pm 0.7$ & 0.2 & $1380^{+530}_{-380}$\\
V392 Hya & K5 or K6 & $18.9 \pm 0.2$ & $0.60 \pm 0.05$ & $0.78 \pm 0.03$
& $6.0 \pm 0.2$ & 0.2 & $3470^{+510}_{-450}$\\
RX 1951 & K7.5 or M0.5 & $17.6 \pm 0.2$ & $0.6 \pm 0.3$\tablenotemark{c} 
& $0.99 \pm 0.18$ & $7.4 \pm 0.6$ & 0.8 & $760^{+240}_{-180}$\\
\enddata
\tablenotetext{a} {Note carefully that these masses are not measured,
but are estimates guided by the models of \citet{bk00}.  They are 
used {\it only} to constrain $R_2$, which depends only on the cube
root of $M_2$, so this does not contribute substantially to the error
budget.}
\tablenotetext{b} {Absolute visual magnitude inferred for the secondary
alone, on the basis of surface brightness and Roche lobe size (see
text.)}
\tablenotetext{c} {The period is above the limit of evolutionary 
scenarios computed by \citet{bk00}, so a conservative estimate of the 
mass is made.}
\label{tab:inferences}
\end{deluxetable}

\clearpage

\clearpage

\clearpage

\end{document}